\documentclass[
aps,
prmaterials,
superscriptaddress,
amsmath,
amssymb,
two column,
showkeys
]{revtex4-2}
\usepackage{graphicx}

\begin{document}

\title{Charge transfer due to defects in hexagonal boron nitride/graphene heterostructures: an \textit{ab initio} study}
\author{Madhava Krishna Prasad}
\affiliation{School of Mathematics, Statistics and Physics, Newcastle University, Newcastle upon Tyne, NE1 7RU, United Kingdom}
\affiliation{Joint Quantum Centre Durham-Newcastle, United Kingdom}
\author{Oras A. Al-Ani}
\affiliation{School of Mathematics, Statistics and Physics, Newcastle University, Newcastle upon Tyne, NE1 7RU, United Kingdom}
\affiliation{Electrical Engineering Technical College, Middle Technical University, Baghdad, Iraq}
\author{Jonathan P. Goss}
\email{jonathan.goss@newcastle.ac.uk}
\affiliation{School of Mathematics, Statistics and Physics, Newcastle University, Newcastle upon Tyne, NE1 7RU, United Kingdom}
\author{Jonathan D. Mar}
\email{jonathan.mar@newcastle.ac.uk}
\affiliation{School of Mathematics, Statistics and Physics, Newcastle University, Newcastle upon Tyne, NE1 7RU, United Kingdom}
\affiliation{Joint Quantum Centre Durham-Newcastle, United Kingdom}
\date{\today}


\begin{abstract}

Using density functional theory (DFT), we study charge transfer between hexagonal boron nitride (h-BN) point defects and graphene in h-BN/graphene heterostructures for a range of intrinsic defects --- nitrogen vacancy, boron vacancy, nitrogen antisite and boron antisite. We show that traditional methods that calculate charge transfer by spatial discrimination of charge to different atoms suffer from the misallocation of charge and introduce an alternative method that relies on the integration of the density of states. We also show that DFT calculations of charge transfer have cell size dependencies due to a change in the density of states in the vicinity of the defect levels. Our results indicate that the nitrogen and boron anitsites do not participate in charge transfer, whereas the nitrogen and boron vacancies experience the transfer of a whole electron. Additionally, we show that a change in the geometry of a defect corresponds to a change in the charge state of the defect. The results of our study will be invaluable for a wide variety of device applications that involve charge transfer between h-BN defects and graphene in h-BN/graphene heterostructures, while our methodology can be feasibly extended to a wide range of point defects and heterostructures.

\end{abstract}



\maketitle



\section{\label{sec:level 1} Introduction}

As the first 2D van der Waals material to be realised, graphene has been the focus of an intense research effort due to its extraordinary properties, impacting a wide range of applications in electronics, sensing, medicine and energy \citep{Wolf2014,chang2010graphene,li2019intrinsic,olabi2021application,pumera2011graphene,pumera2011biosensing,choi2010synthesis}. As a natural complementary material to graphene, hexagonal boron nitride (h-BN) has been combined with graphene to form van der Waals heterostructures, leading to a variety of novel device physics and applications. Examples include graphene devices with very high mobility and very low carrier inhomogeneity \citep{palacios2018atomically,yankowitz2019van,dean2010boron}, graphene spintronic devices with long spin relaxation times and efficient spin injection using h-BN as a substrate/encapsulation layer or tunnel barrier \citep{guimaraes2014controlling,gurram2018electrical,kamalakar2014enhanced}, and graphene field-effect transistors and twistronic devices where h-BN is used to modify the band structure of graphene \citep{yankowitz2012emergence,giovannetti2007substrate}, to name just a few.

In many of these device applications which employ h-BN/graphene (h-BN/Gr) heterostructures, charge transfer involving defects that are inevitably present in h-BN is a critical factor in device performance and operation. For example, charge transfer resulting in the creation of charged traps in h-BN may act as Coulombic scattering centers, lowering carrier mobilities and spin relaxation times in graphene \citep{gosling2021universal,chandni2016signatures}. Using charge transfer in h-BN defects as a resource, spin-dependent tunnelling in magnetic tunnel junctions has been enhanced due to resonant tunnelling through magnetic defect states in an intermediate h-BN layer \citep{chandni2016signatures,asshoff2018magnetoresistance}. Additionally, charge transfer has been used to spectrally and spatially quench single-photon emission from h-BN defects when deposited on functionalized \citep{xu2020charge} and patterned \citep{stewart2021quantum} graphene, respectively. Therefore, given the key role that it plays in a wide range of device applications, a detailed theoretical study of charge transfer involving h-BN point defects in h-BN/Gr heterostructures is of fundamental importance. However, such a detailed theoretical study is still lacking in the literature.

Here, we use density functional theory (DFT) to study charge transfer between h-BN point defects and graphene in h-BN/Gr heterostructures for a range of intrinsic defects. Traditional methodologies of determining the degree of charge transfer involve integration of the charge density distribution. However, we show that such methods suffer from errors due to the misallocation of charge, since no principled way of allocating charge to an atom exists. We therefore propose an alternative methodology of quantifying the degree of charge transfer that circumvents this issue by using the method of integration of the density of states (DoS).  We also show that DFT calculations of charge transfer have cell size dependencies due to a change in the density of states in the vicinity of the defect levels. Along with supporting calculations of the ionisation energies of defects with respect to the work function of graphene, as well as calculations of the band structure and the total charge in each layer of the h-BN/Gr heterostructure, we determine the propensity of charge transfer for the nitrogen vacancy, boron vacancy, nitrogen antisite and boron antisite. Our findings show that the nitrogen and boron antisites do not participate in charge transfer, whereas the nitrogen and boron vacancies experience the transfer of a whole electron. We also show that a change in the geometry of a defect in an h-BN/Gr heterostructure is consistent with a change in the charge state of the defect.


\section{Methodology}\label{sec:method}

Our DFT calculations were performed using the Ab Initio Modelling PROgram\citep{jones1998identification} (AIMPRO) with periodic boundary conditions and the PBE-GGA exchange-correlation functional \citep{perdew1996generalized}. 

Atoms are modelled using norm-conserving separable pseudo potentials~\citep{hartwigsen-PRB-58-3641}, with $1s$-states of B, C and N part of the core.

Kohn-Sham eigenfunctions are represented with a basis of sets of independent $s$- and $p$-Gaussian orbitals with four different exponents centered on atomic sites~\citep{goss-TAP-104-69}, with the addition of one (two) sets of $d$-Gaussian functions for C (B and N) atoms to account for polarization. This amounts to 18 independent Gaussian functions per C atom in the basis, and 28 per B and N atom. Additional sets of functions are located in the vacuum regions to ensure accurate representation of the evanescence.
The charge density is Fourier transformed using plane waves with an energy cutoff of 300\,Ha, leading to energies converged to better than 1{\,meV} with respect to this parameter. 

The Brillouin zone of the primitive structures were sampled using a $16\times16$ $k$-point grid and the Monkhorst-Pack scheme \citep{monkhorst1976special}. Non-primitive cells are modeled using grids with a comparable or denser reciprocal space density.

Structures were optimized by the conjugate-gradient method until the total energy changed by less than $10^{-5}$\,Ha, and forces are less than $10^{-4}$\,a.u. 

The spacing between monolayers was set to 30\,a.u.\ (15.89\,\r{A}), which is approximately five and four interlayer spacings of bulk h-BN~\citep{wang2017graphene} for monolayer and heterostructure models, respectively. 

The optimised in-plane lattice constant of monolayer h-BN was calculated to be 2.514\,\r{A}, in good agreement with previous comparable calculations and with the experimental value of 2.504\,\r{A} \citep{cai2020first,xu2015lattice,janotti2001first,topsakal2009first}.  Similarly, our calculated value of 2.47\,{\AA} for the lattice constant of monolayer graphene is in excellent agreement with the literature~\citep{wang2017graphene}. 
The reproduction of the geometric parameters and band structures (band-structure data are presented in Section\,\ref{sec:results}) of these monolayer systems provides confirmation that the basis sets, sampling and treatment of vacuum are sufficiently accurate to provide confidence in the calculated properties of the more complex systems at the center of this study.

Van der Waals interactions were represented using the Grimme-D3 scheme~\citep{grimme2011density}. As monolayer h-BN and graphene have different in-plane lattice constants, a decision regarding the treatment of the lattice-constants for heterostructures needed to be made.  We have adopted the approach of using a fixed value obtained from the optimisation of the in-plane lattice constant for the combined system.  This lies between the values of the two isolated systems at 2.49\,{\AA}, representing 1\% compressive and tensile strains for h-BN and graphene, respectively. 

The formation energy of a defect, $X$, in a specific charged state, $q$, is 
  
$$    E_f(X,q) = E_\text{tot}(X,q) - E_\text{host} - \sum_i n_i \mu_i + q\left(\varepsilon_\text{VBM} + \varepsilon_F \right),$$

where $E_\text{tot}(X,q)$ is the total energy of the defective supercell of h-BN, $E_\text{host}$ is the total energy of pristine monolayer h-BN of the same size, $n_i$ is the change in the number of atoms of species $i$ relative to pure h-BN and $\mu_i$ is the chemical potential of the species $i$. The formation energies were calculated in the N-rich condition specified by $\mu_\text{N} +\mu_\text{B} = \mu_\text{h-BN},$ where $\mu_\text{N}$ is half the total energy of an N$_2$ molecule and  $\mu_\text{h-BN}$ is the energy per formula unit of monolayer h-BN. $\varepsilon_\text{VBM}$ and $\varepsilon_F$ are the energy of the valence band maximum (VBM) of the host and the electron chemical potential, respectively. 

We have adopted the standard $(q/q')$ notation to denote the charge transition level (CTL) between charges $q$ and $q'$ relative to the VBM.

Periodic boundary conditions, especially for polarised and charged systems, lead to well known systematic errors.
Additionally, charged defects in an anisotropic system, like monolayer h-BN, are electrostatically screened in-plane but unscreened across the vacuum  \citep{komsa2013finite,komsa2014charged}. Correction techniques generally involve extrapolating properties to the dilute limit using data from a range of cell sizes \citep{castleton2004finite,komsa2012finite}.  We have adopted the uniform scaling of the cell sizes of Refs.\,\onlinecite{komsa2014charged,komsa2018erratum},
leading to an uncertainty of the order of $\pm0.1$\,eV in the formation energy \citep{komsa2018erratum}. As we consider defects with low charged states, the maximum charge being $|2e|$, we find that the uncertainty in the valence band position due to the artifical electrostatic field from the charged defect is negligible.

Finally, total electron spin was included as a free parameter during self-consistency and optimisation of defect-containing heterostructures, with the total spin reflecting the population of the spin channels based upon Fermi-Dirac statistics with spin-up and spin-down channels having the same, self-consistent electron chemical potential. Non-integer spins are found in many cases of small supercells, reflecting the partial charge transfer. The role of the band occupation including spin is explored in Section~\ref{sec:discussion}. 

Quantification of charge transfer has been approached in two ways.  
The first is by integrating the DoS of the heterostructure from the Fermi level to the small band gap induced by the formation of the heterostructure. For simplicity, we have performed the integration over the majority spin DoS to obtain the population of free carriers in graphene. This region reflects the charge depleted (transferred) from (to) the Dirac cone. The second involves allocating charge density into h-BN and graphene components by dividing the space between the two materials according to the location of the minimum in the average planar charge-density. 

\begin{figure*}[!ht]
    \includegraphics[]{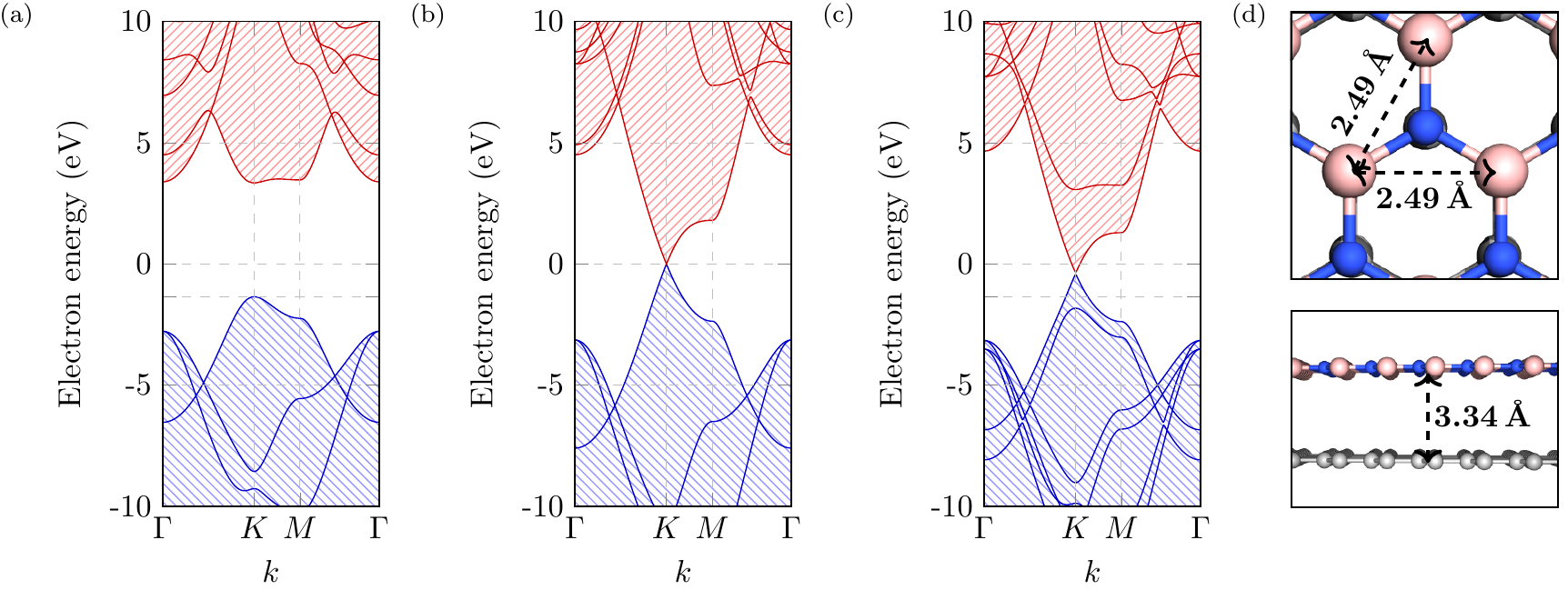}
    \caption{Calculated band structures in the vicinity of the Fermi energy along high-symmetry branches of the Brillouin zone for (a) monolayer h-BN, (b) monolayer graphene and (c) an h-BN/graphene heterostructure. Blue and red lines represent nominally occupied and empty bands, respectively, with the underlying shading highlighting the envelopes of the valence and conduction bands. The zero on the energy scale is the Dirac-point in pristine graphene, with the other systems aligned so their vacuum levels coincide.  (d) Structure of the h-BN/graphene heterostructure, annotated with relevant lengths.  Blue, pink and grey spheres represent N, B and C atoms, respectively.}
    \label{fig:pristine}
\end{figure*}

Then the integrated charge density in each half is allocated to h-BN or graphene, as appropriate.  A uniform mesh density which was sufficient to converge the total charge in the supercell to $10^{-2}e$ was used.

The net charge in each volume is the difference between the integrated electron density and the ion charges and the degree of charge transfer is the difference in the total charge of each layer from the monolayer case. 

The degree of charge transfer quoted has been converged with cell size to two decimal places. As we shall show, we find that the degree of charge transfer converges with cell size significantly faster using the integration of DoS than integration of charge density. We explore the dependence of the degree of charge transfer on cell size in Section~\ref{sec:discussion}. 

\section{Results}\label{sec:results}

\subsection{Pristine h-BN, graphene and h-BN/graphene heterostructure}

The calculated band structure of pristine h-BN is shown in Fig.\,\ref{fig:pristine}a, which shows a band gap of 4.6\,eV, in agreement with comparable calculations  \citep{topsakal2009first}.  This is an underestimate compared to the experimental value of 6.1\,eV \citep{elias2019direct}, which is a well-known effect of DFT-PBE calculations \citep{cohen2008insights}, but we note the valence band dispersion is consistent both with comparable modelling and angle-resolved photoemission spectroscopy measurements \citep{henck2017direct,topsakal2009first,wickramaratne2018monolayer}. The ionisation energy of h-BN was found to be 5.9\,eV, also in agreement with comparable calculations \citep{liu2020extrapolated}. 

The calculated band structure of graphene, shown in Fig.\,\ref{fig:pristine}b, exhibits the Dirac cone at the $K$-point in accordance with experiment \citep{sprinkle2009first}. The work function of graphene was calculated to be 4.3\,eV, comparable to the experimental value of 4.6\,eV and other PBE-GGA calculations \citep{yu2009tuning,ziegler2011variations}. 

Fig.\,\ref{fig:pristine}c shows the h-BN/Gr band structure, showing it can be understood as a simple superposition of the band structures of the individual layers. However, a $\sim0.1$\,eV band gap opens up near the Dirac point, in agreement with existing literature \citep{torres2022band}.

The baseline degree of charge transfer in the pristine heterostructure was negligible. The deviation from zero for the defective cases indicates charge transfer. We can now proceed to the analysis of the defective cases of isolated h-BN and h-BN/Gr heterostructure.


\subsection{The nitrogen vacancy, V\textsubscript{N}}

\begin{figure*}[!htbp]
\centering
    \includegraphics[]{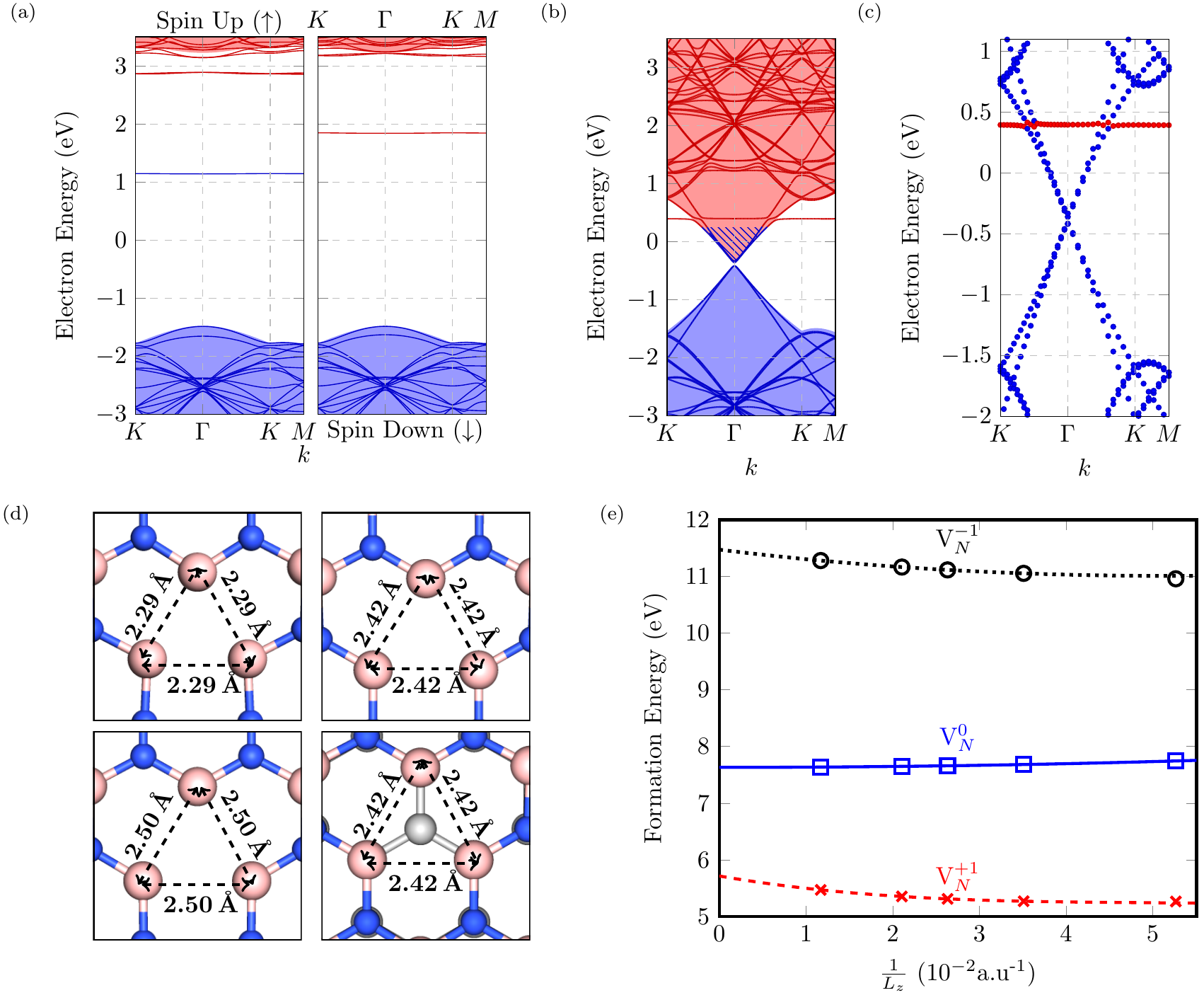}
    \caption{Band structures of V$_\text{N}^0$ in (a) h-BN, with the underlying shading corresponding to the occupied and empty bands on pristine h-BN, and (b) h-BN/Gr. The hatched shading in (b) indicates the filling of the graphene bands up to the Fermi level, with the underlying shading indicating occupied and empty bands of the corresponding defect-free h-BN/Gr for comparison.  (c) Localization of the bands to h-BN (red) or graphene (blue) based upon Mulliken populations. Colours and scales are as in Fig.\,\ref{fig:pristine}. (d) Plan-view schematics of V$_\text{N}^0$ (top left) and V$_\text{N}^{+1}$ (bottom left) in isolated h-BN, V$_\text{N}^{+1}$ in monolayer h-BN with the in-plane lattice constant of that of the heterostructure (top right), and V$_\text{N}$ in h-BN/Gr heterostructure (bottom right). (e) Plot of formation energy as a function of cell size (points) with cubic polynomial fits (lines). }
    \label{fig:vn in h-BN/Gr}
\end{figure*}

The removal of a single nitrogen atom results in a nitrogen vacancy. V\textsubscript{N} has been optimised in an h-BN monolayer in several charge states and cell sizes. We find that $\text{V}_\text{N}^{+1}$, $\text{V}_\text{N}^{-1}$ and $\text{V}_\text{N}^0$ possess $D_{3h}$ symmetry, and favour low spin states, in agreement with literature \citep{huang2012defect,weston2018native,sajid2018defect}. As shown in the band structure (Fig.\,\ref{fig:vn in h-BN/Gr})a, V\textsubscript{N} leads to three gap states. In the spin-up channel, there is a degenerate unoccupied state close to the conduction band and a singly-occupied non-degenerate level $\sim$2.5\,eV above $\varepsilon_\text{VBM}$.  In the heterostructure, the corresponding  gap-level is depopulated.  The heterostructure system favours a singlet state, corresponding to $\text{V}_\text{N}^+$ and a non-magnetic configuration of a partially occupied Dirac cone (Fig.\,\ref{fig:vn in h-BN/Gr}b). To further illustrate the association of the bands near the Fermi energy with the defect, Fig.~\ref{fig:vn in h-BN/Gr}c shows the same band structure where each state is denoted as h-BN or graphene based upon Mulliken populations: red circles indicate bands more localised in the h-BN layer and blue circles indicate bands more localised in the graphene layer. Therefore, it is clear that the defect level is localised in h-BN. 

The extrapolation of the formation energies of different charge states of V\textsubscript{N} to the infinitely dilute solution limit is shown in Fig.\,\ref{fig:vn in h-BN/Gr}e.  The range of cell sizes used for extrapolation are, $4a\,\times\,4a, 6a\,\times\,6a, 8a\,\times\,8a, 10a\,\times\,10a$ and  $18a\,\times18a$. We obtain an extrapolated value of $E_f(\text{V\textsubscript{N}},0)=7.6\,\text{eV}$, which agrees well with the literature value of 7.7\,eV \citep{huang2012defect}. The cell sizes used in the extrapolation of the formation energies are consistent across all defects in this paper. The calculated (0/+) level is 1.9\,eV, placing it 4.0\,eV below vacuum.  ($-$/0) lies at 3.9\,eV, which is 2.0\,eV below vacuum. Both levels lie above the work-function of graphene, and are in good agreement with previous calculations~\citep{wu2017first,liu2020extrapolated}.
Energetically, the location of the (0/+) level suggests electron transfer to the graphene layer should occur, consistent with the band structure and spin state. 
Furthermore, the calculated net charge of the defective h-BN is $+e$ in the heterostructure, indicating that a whole electron was transferred to the graphene layer. The structure of the defect in the heterostructure is similar to V$_\text{N}^{+1}$ in isolated h-BN (Fig.\,\ref{fig:vn in h-BN/Gr}d), consistent with charge transfer.

\begin{figure*}[!ht]
    \includegraphics[]{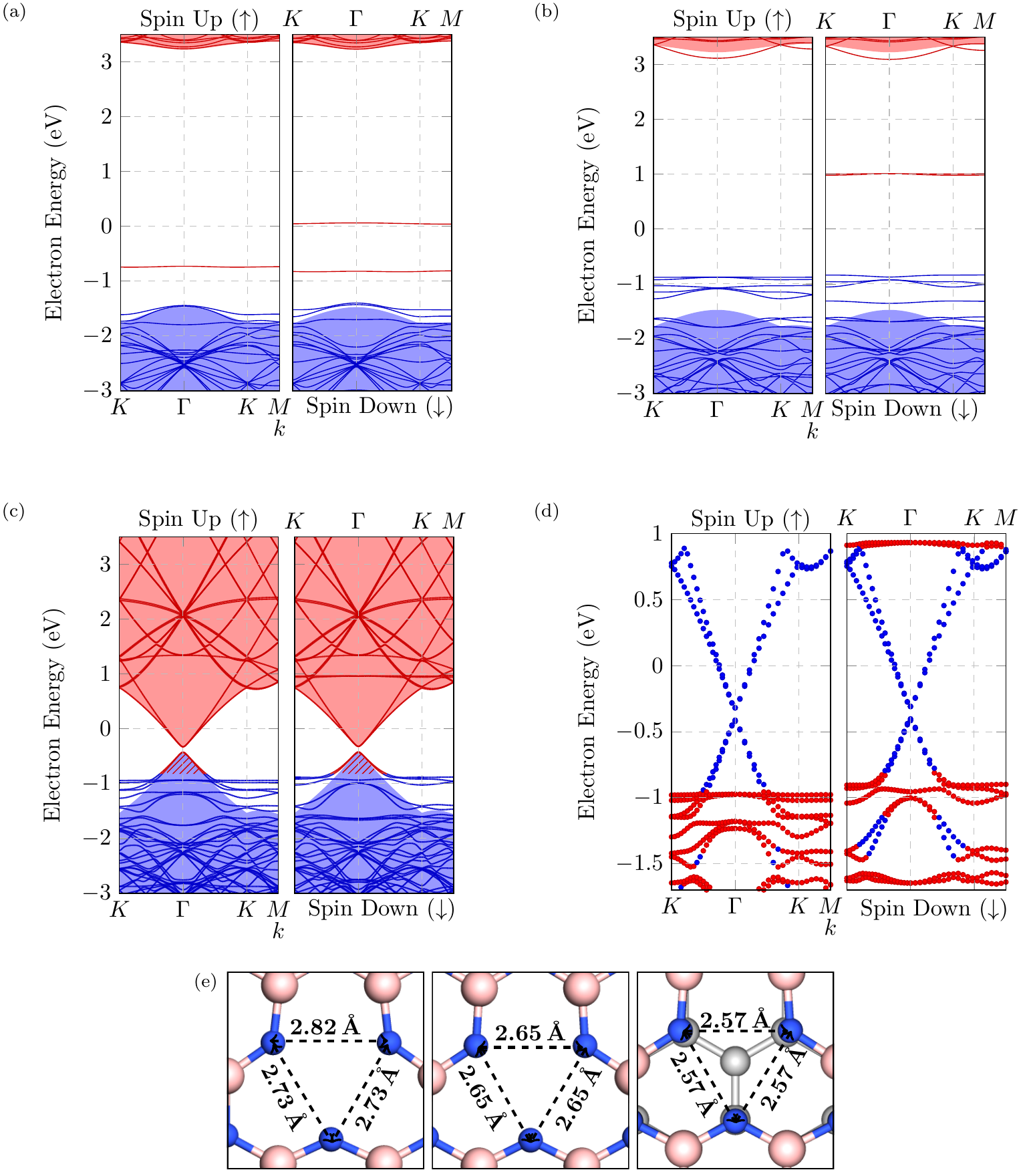}
\caption{Band structures of (a) V$_\text{B}^0$ in h-BN, (b) V$_\text{B}^{-1}$ in h-BN, and (c) V\textsubscript{B} in h-BN/Gr.  {(d)} Mulliken populations analysis. (e) From left to right: schematics of V$_\text{B}^0$ and V$_\text{B}^{-1}$ in  h-BN and V\textsubscript{B} in h-BN/Gr. Colors and scales are as in Fig.\,\ref{fig:pristine} and Fig.\,\ref{fig:vn in h-BN/Gr}c.}
\label{fig:vb in h-BN/Gr}
\end{figure*}

Ionizing V\textsubscript{N} depopulates bonding orbitals leading to neighboring B-atoms relaxing outwards, leading to increased B--B distances (Fig.~\ref{fig:vn in h-BN/Gr}d). The geometry of V$_\text{N}$ in the heterostructure resembles that of V$_\text{N}^{+1}$ in isolated h-BN, consistent with charge transfer.


\subsection{The boron vacancy, V\textsubscript{B}}

In agreement with previous studies, our optimised ground-state structure for $\text{V}_\text{B}^0$ has $C_{2v}$, arising from a Jahn-Teller distortion~\citep{huang2012defect,ivady2020ab}.  V\textsubscript{B} acts as an acceptor~\citep{weston2018native} with the $-1$ charge state found to be a spin-triplet with $D_{3h}$ symmetry, whereas the $-2$ charge state is a doublet with $C_{2v}$ symmetry, in agreement with literature \citep{huang2012defect,weston2018native,strand2019properties}. 

Band structures of V$_\text{B}^0$ and V$_\text{B}^-$ in monolayer h-BN are shown in Fig.\,\ref{fig:vb in h-BN/Gr}. There are defect levels within the band gap in each spin channel in the neutral charge case, which are non-degenerate and unoccupied.  An occupied band in the spin-up channel, corresponding to one of the empty spin-down states lies in the valence band.  In the negative charge state, the higher symmetry leads to a doubly degenerate unoccupied spin-down band deep in the band-gap, and an occupied degenerate state close to $\varepsilon_\text{VBM}$ that mixes with the valence band states, resulting in a multitude of defect related bands around this energy.

Previous studies~\citep{weston2018native} indicate V\textsubscript{B} is a triple acceptor, and we find single and double acceptor levels at 1.0\,eV (4.9\,eV below vacuum) and 5\,eV \citep{weston2018native,liu2020extrapolated}. The triple acceptor level is very close to the conduction band.

\begin{figure*}[!ht]
\centering
    \includegraphics[]{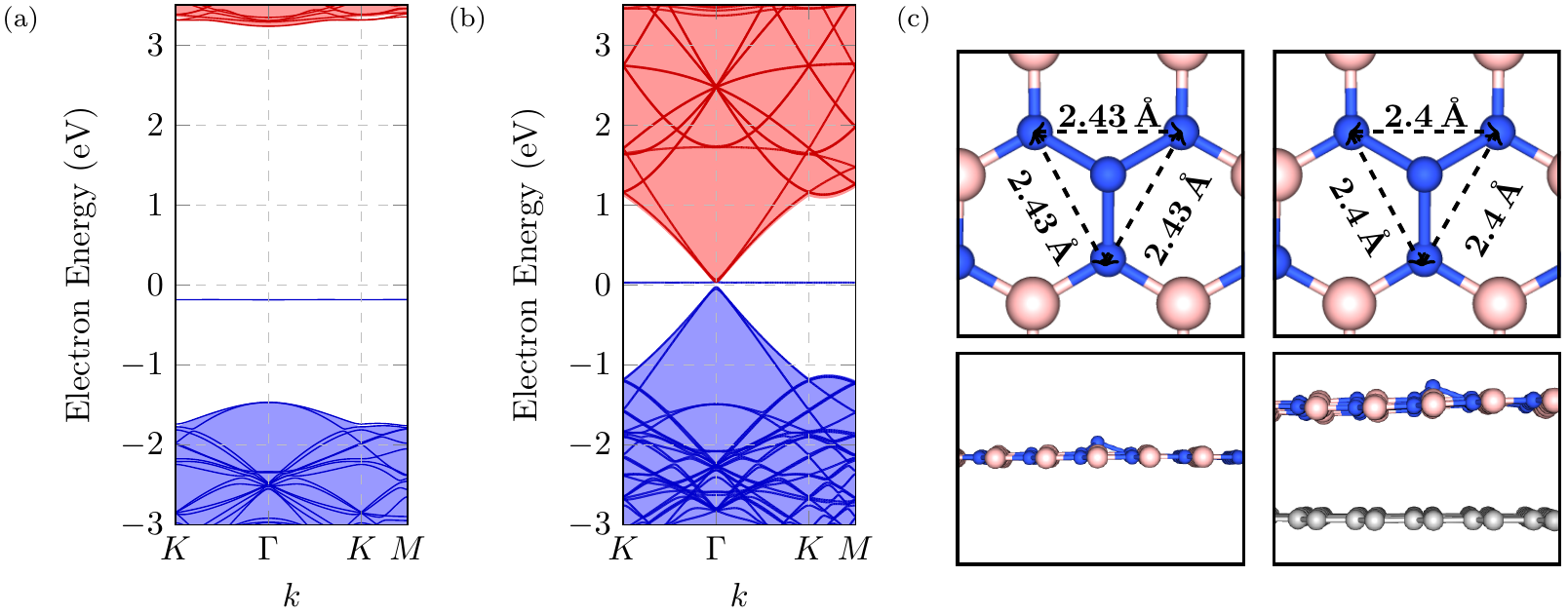}
    \caption{Band structures of  N$_\text{B}^0$ in (a) h-BN and (b) h-BN/Gr.  (c) Schematic representations of the plan and side views of the corresponding structures, showing the displacement of the antisite nitrogen from the h-BN plane.   Colors and scales are as in Fig.\,\ref{fig:pristine}.}
    \label{fig:nb in hbn/gr}
\end{figure*}

As the calculated ($-$/0) level of V$_\text{B}$ in pristine h-BN is below the work function of graphene, it is thermodynamically favourable for an electron to be transferred from graphene to h-BN.

Figure \ref{fig:vb in h-BN/Gr}c shows the band structure of $\text{V}_\text{B}$ in the heterostructure.  The similarity of this band structure to that of  V$_\text{B}^{-1}$ (Fig.~\ref{fig:vb in h-BN/Gr}b) strongly indicates a change in the charge and spin state of the defect. The localisation of bands (Fig.~\ref{fig:vb in h-BN/Gr}d) confirms the association of the relevant bands to the h-BN, as does the equilibrium geometry of the heterostructure being close to that of the negative charge state in monolayer h-BN, Fig.~\ref{fig:vb in h-BN/Gr}e.

Additionally, calculation of the total charge for each layer confirms the transfer of a whole electron and the magnetic moment of the defect was found to be $2\mu_\text{B}$. We note that this is significantly larger than the degree of charge transfer and magnetic moment found in Ref.\,\onlinecite{park2014interlayer}. 

\subsection{The nitrogen antisite, N\textsubscript{B}}

The replacement of a boron atom by a nitrogen atom results in the nitrogen antisite, N\textsubscript{B}. We find that in monolayer h-BN this centre favours a spin-doublet in its uncharged state and a singlet in the positive charge state. Neutral N\textsubscript{B} possesses an occupied non-degenerate level deep within the band gap, as seen in Fig.~\ref{fig:nb in hbn/gr}a.  The antisite nitrogen atom moves out-of-plane (Fig.~\ref{fig:nb in hbn/gr}c) resulting in $C_{3v}$ symmetry, but this does not happen to the positively ionised case which we find to be planar ($D_{3h}$ symmetry).

In h-BN/Gr the band structure (Fig.~\ref{fig:nb in hbn/gr}b) shows the occupied defect band to lie in the band gap, and the antisite nitrogen atom moves out of the h-BN plane.  This is consistent with the electrical levels determined for the antisite: the (0/+) level of N\textsubscript{B} is calculated at 0.9\,eV, which is 5.0\,eV below the vacuum, in agreement with literature\citep{liu2020extrapolated}, and hence the ionisation energy of the defect exceeds the work function of graphene. Thus it is energetically unfavourable for this defect to donate any charge to the neighbouring graphene. Indeed, no change in the total charge was calculated in each layer. We therefore conclude that under equilibrium conditions N\textsubscript{B} would not donate or accept charge with graphene.


\subsection{The boron antisite, B\textsubscript{N}}

\begin{figure*}[!htbp]
\centering
    \includegraphics[]{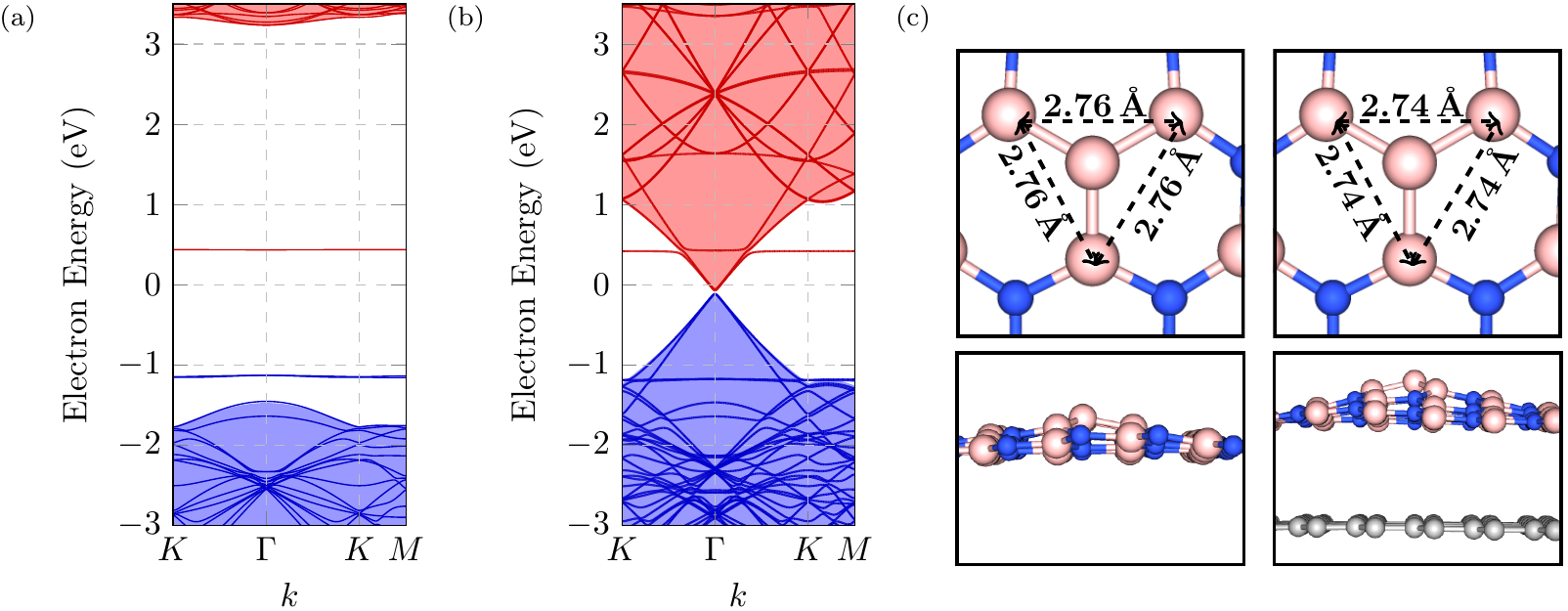}
    \caption{Band structures of  B$_\text{N}^0$ in (a) h-BN and (b) h-BN/Gr.  (c) Schematic representations of the plan and side views of the corresponding structures, showing the displacement of the antisite boron from the h-BN plane.   Colours and scales are as in Fig.\,\ref{fig:pristine}.}
    \label{fig:bn in h-BN/Gr}
\end{figure*}

Finally, we summarize the results for the boron antisite, B\textsubscript{N}. Like its nitrogen counterpart we obtain a spin singlet ground state in its neutral charge state and a spin doublet in its ionised state.  The introduction of the defect into h-BN leads to three states in the band gap.  A doubly-degenerate band lies close to $\varepsilon_\text{VBM}$, and a non-degenerate unoccupied band lies mid-gap (Fig.~\ref{fig:bn in h-BN/Gr}a). 

In h-BN/Gr the occupied states lies below the band gap and the empty state above, so the band structure indicates charge transfer to be unlikely.  Furthermore, ($-$/0) for B\textsubscript{N} is calculated to be 2.8\,eV, i.e{.} 3.1\,eV below vacuum, placing the acceptor level well above the work function of graphene. These values of CTLs are consistent with literature \citep{weston2018native,liu2020extrapolated}, and the lack of charge transfer is confirmed by the integrated charge density showing negligible change in total charges on the two layers.  

\section{Discussion\label{sec:discussion}}

\begin{figure*}[!htbp]
    \centering\includegraphics[]{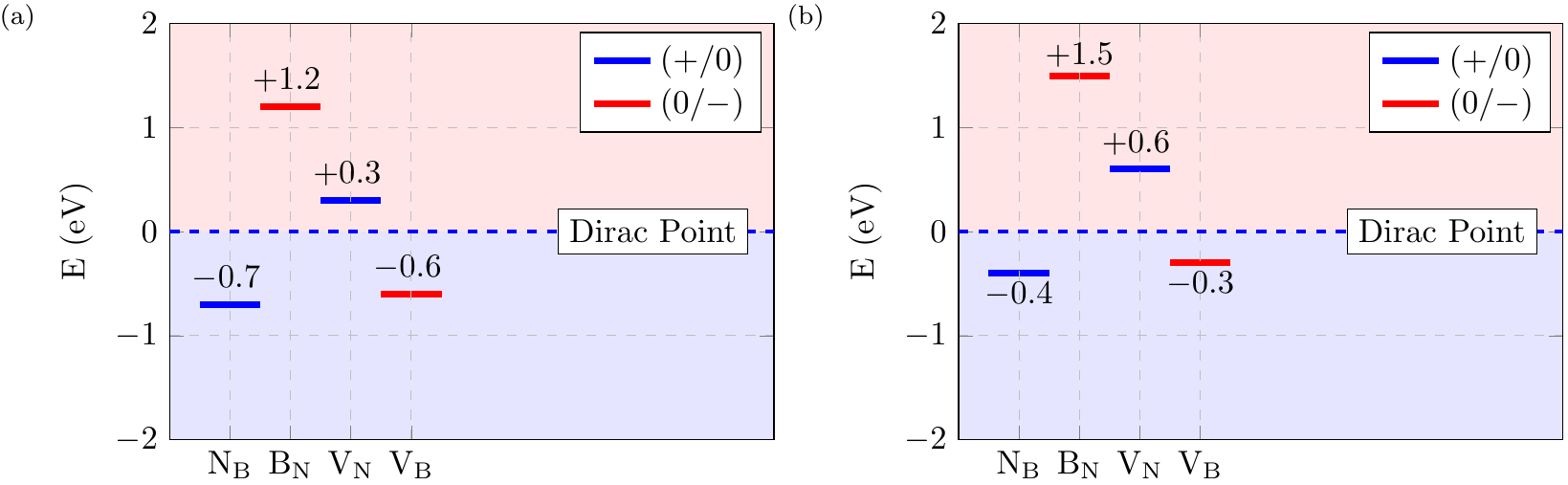}
    \caption{Charge transition levels of defects studied in this paper, calculated using (a) PBE-GGA functional, relative to the calculated work function of graphene, and (b) HSE functional, relative to the experimental work function of graphene.}
     \label{fig: ctls}
\end{figure*}

\begin{figure*}[!htbp]
    \centering\includegraphics[]{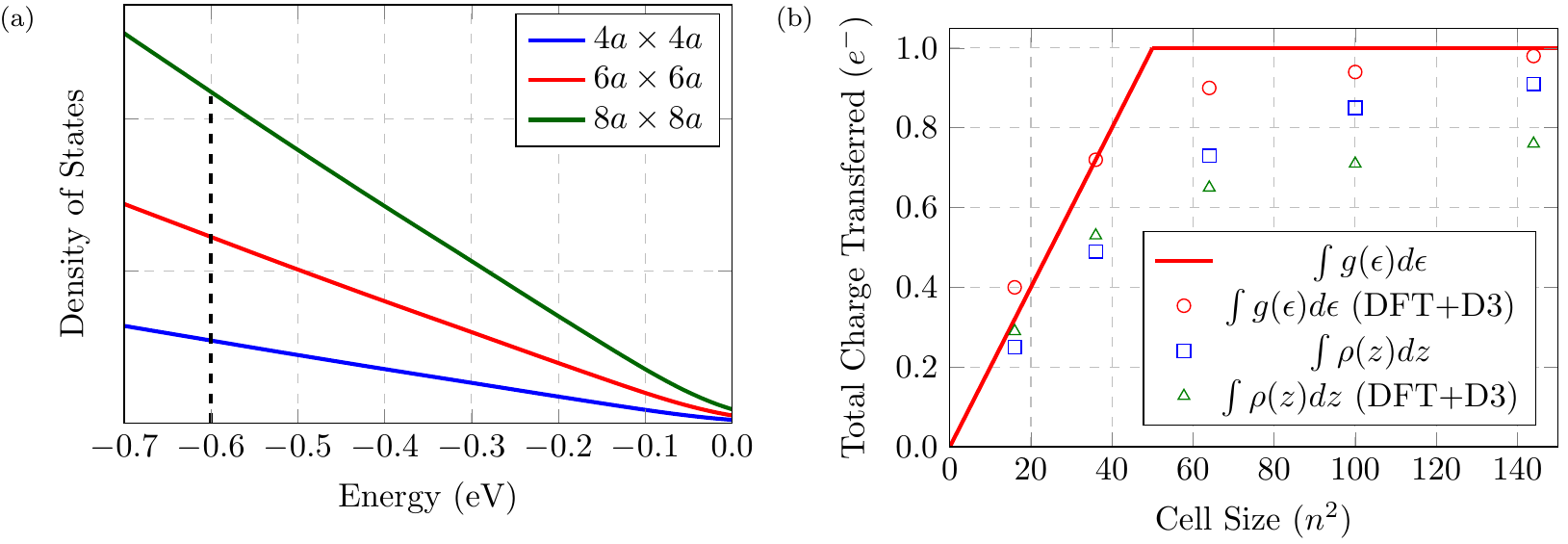}
    \caption{(a) A plot of the total electronic DoS for pristine graphene at the approximately linear region close to the Dirac point. The vertical dotted black line is the ($-$/0) level of V\textsubscript{B} in isolated h-BN relative to the Dirac point in pristine graphene. (b) The degree of charge transfer obtained by the integration of the DoS of graphene and the heterostructure (red line and red circles, respectively) and by the charge density distribution with and without van der Waals forces (blue squares and green triangles, respectively).}
    \label{fig: dos scaling}
\end{figure*}

\begin{figure}[!htbp]
    \centering
    \includegraphics[]{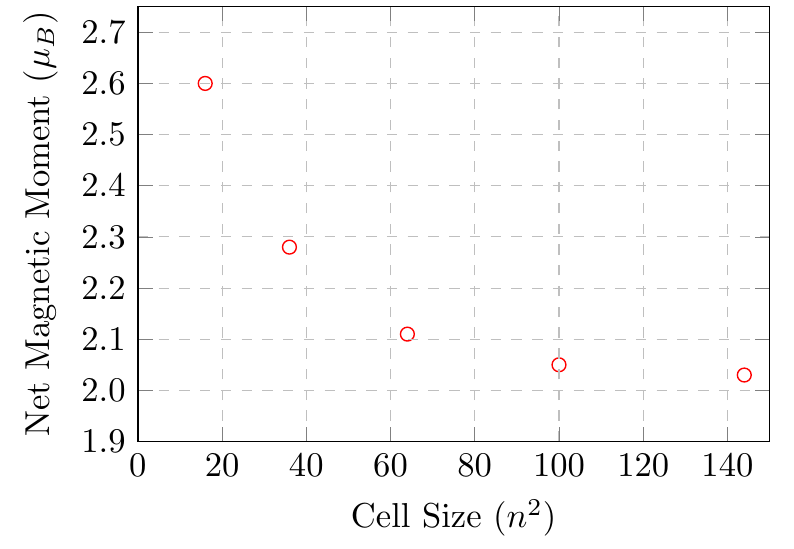}
    \caption{A plot of $\mu_B$ with respect to the cell size for V\textsubscript{B} in h-BN/Gr.}
    \label{fig: mu_B scaling}
\end{figure}

It is informative to compare charge transfer across the four primary native defects studied. The CTLs of the defect in isolated h-BN with respect to the Dirac point of graphene is a good predictor of the propensity of charge transfer. The CTLs of the antisites are such that there is an energy cost for charge transfer to occur, whereas for vacancies it is thermodynamically favorable for charge transfer to occur as the donor (acceptor) state lies above (below) the Dirac point in graphene (Fig.\,\ref{fig: ctls}). 

It is also instructive to reflect upon potential impact of the choice of exchange-correlation functional. CTLs of native defects in h-BN obtained using screened-exchange methods (HSE) can be estimated from PBE-GGA values \citep{liu2020extrapolated}.  In Fig.\,\ref{fig: ctls} PBE-based CTLs calculated in this paper and the HSE-based CTLs obtained from Ref.\,\onlinecite{liu2020extrapolated} are plotted, where the values are stated relative to the work function of graphene. Differences in the location of CTLs between PBE-GGA and HSE estimates have been shown to be largely systematic~\citep{liu2020extrapolated} and whether donor and acceptor levels lie above or below the Dirac-point is independent of approach. Hence, computation of the propensity for charge transfer between defects in h-BN and graphene can be performed with PBE-GGA functionals to take advantage of the relatively lower computational cost.

We now turn to the key impact of simulation cell-size and method of estimate on the degree of charge transfer. For this we use V\textsubscript{B} as a case study. 

We begin with the data resulting from the integration of the charge density when dividing the volume into two halves based on the plane containing the minimum of the average charge density. Fig.~\ref{fig: dos scaling}b shows the degree of charge transfer for two cases.  In the absence of the van der Waals correction, the inter-plane separation is larger (4.2\,\r{A}) than with the correction (3.3\,\r{A}) and for these data the average charge density between the graphene and h-BN drops to a very low value.  When the van der Waals correction is included, the overlap in the charge density coming from the two materials is much greater, and the minimum value of the charge density between the layers is much greater.  
In the absence of the van der Waals correction, the integration of the charge density suggests that the transfer of a whole electron would be expected, with the trend in the data suggesting the integrated charge asymptotically approaches one, whereas for the corrected case the convergence is to a much smaller quantity.

From a fundamental physics point of view, there is no principled way to spatially allocate electron charge to a specific atom, and in this case to either h-BN or graphene.  For the cases with different inter-plane distances there is a difference in the evanescent drop of charge density, and charge density allocated using proximity suggests that the degree of charge transfer is strongly dependent upon the inter-plane distance.

We now turn to the evaluation of charge transfer based upon the electronic DoS. The use of the electronic DoS is distinct from integration of charge density, as it takes into account the separation in energy of bands associated with graphene and h-BN. 
As the CTLs of the point defects examined in this paper lie within a linear regime of the graphene DoS, $g(\epsilon)$, we approximated the graphene DoS as $g(\epsilon)=n^2\lambda\epsilon$, where $n$ is the number of lattice constants in the supercell and $\lambda$ is the gradient of the primitive pristine graphene DoS, found to be $0.055\,\text{eV}^{-2}$. Then, to estimate the cell size required to observe a charge transfer of $N$ electrons, we take a fixed value of the location of the defect CTL and require the graphene DoS between this level and the Dirac-point to account for one charge carrier.  The integrated DoS is determined as 
\begin{equation}
    \int_{0}^{\mu_\text{CTL}} n^2\lambda \epsilon d\epsilon = N\quad\Rightarrow \quad
    n =\frac{1}{2\mu_\text{CTL}}\sqrt{\frac{N}{\lambda}}.
\end{equation}
Here we have taken the Dirac-point to be at zero on the electron energy scale. Then, for single electron or hole transfer
\begin{equation}
    n=\left|\frac{1}{\mu_{\text{CTL}}\sqrt{\lambda}}\right|. \label{eq:min}
\end{equation}

For V\textsubscript{B} and for $\mu_{\rm CTL}$ located 0.6\,eV from the Dirac point, the minimum cell size needed to observe a whole electron transfer would be approximately 50 times larger than the primitive. Fig.\,\ref{fig: dos scaling}a shows the scaling of the density of states near the Dirac point for different cell sizes, illustrating that the integrated DoS between the CTL and the Dirac-point increases with cell size. It also shows that there is a minimum cell for which the area under the graphene DoS is sufficient to allow for a whole electron transfer.  The model in Eq.~\ref{eq:min} is highly simplified and does not account for the self-consistent variation in the location of the defect band with changes in occupation or the dispersion in the defect band.  Fig.~\ref{fig: dos scaling}b shows that the estimate for the minimum size in Eq.~\ref{eq:min} is significantly smaller than that implied by the calculated charge transfer from the integrated charge density. 

The DoS model can be developed further by using the electron energy specta from the heterostructures.  For the combined systems there is a defect band associated with the point defect that exhibits relatively small amounts of dispersion, and for V\textsubscript{B} this lies below the Dirac-point of the neighboring graphene.  As with the more elementary model DoS approach, as the simulation system size increases the underlying graphene DoS increases and the dispersion in the defect band decreases.
Once the underlying graphene DoS in the vicinity of the localised V\textsubscript{B} band is sufficiently large, the integrated DoS above the defect band exceeds one electron.  Once this happens a whole electron is transferred, filling the localized defect band. Further increases in the simulation system size does not increase the integrated DoS between the Fermi energy and the band gap, as the defect band is filled and there is no empty DoS associated with the h-BN or defect to populate from the graphene DoS in the vicinity of the Dirac point. Indeed, using the DoS estimate we found that a cell size greater than $12a\times12a$ showed a whole electron transfer within computational uncertainty (Fig.~\ref{fig: dos scaling}b), and even cells as small as $12a\times12a$ estimate the transfer to be as much as 98\% of an electron.

Given that the two approaches yield such significant differences in the estimate of the charge transfer, it is important to resolve which approach, if either, produces the more reliable estimate.  To answer this question, we address some properties of the system that are independent of any attempt to separate the charge allocation to graphene or h-BN.

First, if the degree of charge transfer varies with cell size and converges to less than one carrier, as predicted by examination of the spatial variation in the charge density, the total effective electronic spin of these systems would be expected to follow a comparable pattern. 
Comparing the calculated effective electronic spin plotted in Fig.~\ref{fig: mu_B scaling} with the charge transfer estimates in Fig.~\ref{fig: dos scaling}, we see that the degree of charge transfer converges with respect to the cell size at the same rate as the DoS calculations.  The effective spin of V\textsubscript{B} converges rapidly to $S=2$, corresponding to the spin-state of the negatively charged vacancy in isolated h-BN and consistent with a whole electron transfer from the graphene.

Secondly, band structure and analysis of the electronic orbitals of V\textsubscript{B} in h-BN/Gr are consistent with it being in the negative charge state. For example, V\textsubscript{B} experiences a Jahn-Teller distortion from $D_{3h}$ to $C_{2v}$ in the neutral charge state in isolated h-BN, whereas the negatively charged spin-triplet case retains the $D_{3h}$ symmetry.  In our calculations, the cell-size converged result shows a geometry indistinguishable from the $D_{3h}$ symmetry case in isolated h-BN. All the available data, other than the integrated charge density, points to the defect being fully ionized and not to a situation with a partial charge transfer.  This casts some light on the result previously published for charge transfer between graphene and h-BN~\citep{park2014interlayer}, which predicted 50{\%} of an electron transfer and a total effective spin of $S=3/2$.  This result was obtained using a simulation cell which we show in this paper does not yield a converged effective spin. Furthermore, the method adopted to estimate the charge transfer was based upon the charge density rather than the band structure.
\section{Conclusion}

In this paper, we have shown that routinely employed methods of determining charge transfer based on spatially allocating charge density results in the misallocation of charge. This becomes especially important in the case of 2D material heterostructures because charge is distributed in the delocalized  $\pi$-states where the distinction between bands associated with dissimilar materials is primarily in terms of their energy rather than their spatial distribution. We therefore adopted an alternative method based on the integration of the electronic DoS, where for the present application we avoid the error of assigning charge in a spatial location to a plane of atoms by integrating the states which have been depleted (filled) from (in) the donor (acceptor) species. In complete support of this approach, we found that the degree of charge transfer with respect to cell size obtained from the integration of DoS follows closely the convergence of the effective electronic spin in the system -- the magnitude of which is closely related to the population of the localised defect states. The magnetic properties are inconsistent with the estimate of the charge transfer from charge-density integration.

We also draw conclusions in the context of the specific material system we have analyzed.  We have shown that the position of charge transition levels of defects in h-BN with respect to the work function of graphene can be used to predict the propensity for charge transfer. From calculations of the CTLs, band structure and quantity of charge transfer, we conclude that N\textsubscript{B} and B\textsubscript{N} do not undergo charge transfer, whereas V\textsubscript{N} and V\textsubscript{B} exchange a whole electron with graphene.

Our conclusions are supported by a combination of band structure, integrated charge density and geometric changes associated with ionized forms of the vacancies.

Critically, there is a clear dependence of charge transfer with supercell dimensions and there is a need to perform calculations of charge transfer quantification in a sufficiently large cell size to achieve convergence.  This is in part a consequence of the localised nature of the states involved in the defects in h-BN, as well as the delocalised states in the graphene. It can also be understood in terms of the graphene density of states in the vicinity of the donor or acceptor band of the defect in the heterostructure. For V\textsubscript{B}, we found that $12\times12$ unit cells were sufficiently large to approximate the dilute limit and we predict that this should be the case for defects where the defect bands have a similar degree of localisation and an acceptor/donor level with a similar energy difference from the Dirac point.

Finally, it is important to note that the native defects studied here serve as prototypes for a much wider range of point defects in h-BN. The principles presented here apply quite generally and the likelihood for charge transfer to take place can be gauged from a knowledge of the location of the donor or acceptor levels relative to the graphene Dirac-point. The results of our study will be key for the design of a wide range of devices that involve charge transfer in van der Waals heterostructures, such as spin valve devices using magnetic defect states for spin-dependent tunnelling, single-photon emitters with electrical charge control and highly sensitive devices for biosensing applications \citep{asshoff2018magnetoresistance,yu2022electrical,white2022vdw,chang2010graphene}. 


\section{Acknowledgements}

We would like to thank Prof. Patrick Briddon and Dr. Mark Rayson for their support in the usage of AIMPRO for the DFT calculations.


\bibliographystyle{apsrev}

\end{document}